\documentclass[%
 reprint,
 amsmath,amssymb,
 aps,
 pra,
]{revtex4-2}

\usepackage{graphicx}
\usepackage{dcolumn}
\usepackage{bm}
\usepackage{amsmath}
\usepackage{tikz}
\usetikzlibrary{quantikz}
\usepackage{chemformula}
\usepackage[colorlinks=true]{hyperref}
\usepackage{subfigure}
\usepackage{changepage}

\usepackage{ulem} 

\begin{document}

\preprint{APS/123-QED}

\title{TETRIS-ADAPT-VQE: An adaptive algorithm that yields shallower, denser circuit ans\"atze}

\author{Panagiotis G. Anastasiou$^1$}
\author{Yanzhu Chen$^1$}
\author{Nicholas J. Mayhall$^2$}
\author{Edwin Barnes$^1$}
\author{Sophia E. Economou$^1$}

\affiliation{$^1$Department of Physics, Virginia Tech, Blacksburg, VA 24061, U.S.A. \\
	$^2$Department of Chemistry, Virginia Tech, Blacksburg, VA 24061, U.S.A.}

\date{\today}

\begin{abstract}
Adaptive quantum variational algorithms are particularly promising for simulating strongly correlated systems on near-term quantum hardware, but they are not yet viable due, in large part, to the severe coherence time limitations on current devices.
In this work, we introduce an algorithm called TETRIS-ADAPT-VQE, which iteratively builds up variational ans\"atze a few operators at a time in a way dictated by the problem being simulated. This algorithm is a modified version of the ADAPT-VQE algorithm in which the one-operator-at-a-time rule is lifted to allow for the addition of multiple operators with disjoint supports in each iteration.
TETRIS-ADAPT-VQE results in denser but significantly shallower circuits, without increasing the number of CNOT gates or variational parameters. Its advantage over the original algorithm in terms of circuit depths increases with the system size. Moreover, the expensive step of measuring the energy gradient with respect to each candidate unitary at each iteration is performed only a fraction of the time compared to ADAPT-VQE.
These improvements bring us closer to the goal of demonstrating a practical quantum advantage on quantum hardware.

\end{abstract}

\maketitle


\section{Introduction}

It is widely hoped that noisy intermediate-scale quantum (NISQ) devices can find applications in the field of quantum simulation, where the archetypal problem is to find the eigenvalues of a given Hamiltonian~\cite{LloydUniversalQsim}. 
To avoid the large resource overhead of the quantum phase estimation algorithm (PEA)~\cite{peaKitaev}, which involves a large number of qubits coherently evolving under very deep circuits~\cite{Cleve1998}, and is thus not expected to offer an advantage over classical simulation in the near future, a hybrid quantum-classical algorithm, the Variational Quantum Eigensolver (VQE) was introduced and experimentally realized for small chemical systems; this algorithm aims to shorten quantum circuits by leveraging the power of classical optimization~\cite{VQEPeruzzo2014}. 

Based on the variational principle, the VQE prepares a parametrized guess wavefunction (known as an ansatz) using a quantum circuit on the quantum processor. It then iteratively optimizes the variational parameters in order to minimize an objective function, usually the energy of the system being simulated. In each iteration, any evaluation of the objective function and/or its gradient is realized by measurements on the quantum processor, and some classical optimization scheme is chosen to update the parameters on the classical processor. 
Since in the VQE the quantum processor is only needed for measuring observables of the ansatz state, the quantum circuits involved are shorter compared to PEA. 
The success of VQEs crucially depends on a good choice of ansatz. On the one hand, the ansatz should be expressive enough so that the ground state of the Hamiltonian can be accurately approximated, and on the other hand it should be special enough so that classical optimization of the variational parameters can be efficiently carried out. 
Finally, the ansatz should also be hardware friendly so that the trial state can be successfully prepared as a quantum circuit on a NISQ processor.

Depending on which of the above properties they are designed to address, the most commonly used ans\"atze can be roughly divided into two families: the hardware efficient~\cite{GanzhornBarkoutsos19, Gard2020, FontanaColes} and the chemically-inspired ans\"atze~\cite{QCSESmartMazzioti,OOUCC, kupGUCC}. The former include circuits consisting of repeating layers of parametrized single-qubit rotation gates and entangling gates, which are easy to implement on a given quantum processor. Such designs take very general forms to ensure expressivity. Nevertheless, it has been shown that a completely problem-agnostic VQE ansatz would hinder the classical optimization~\cite{McClean2018}. 
In contrast, the latter are inspired by classical computational chemistry and consist of much more complicated circuits. 
A widely used ansatz in this category is the unitary extension of Coupled Cluster Singles and Doubles (UCCSD), which forms the basis of many others. 
They are known from classical computational chemistry to possess several desirable features (such as being both variational and size-extensive), but they correspond to relatively deep circuits that are difficult to realize on devices with limited coherence times~\cite{QCCrev}, and their performance is sensitive to the specific Trotterization used in their implementation~\cite{Grimsley2020}.

Among the numerous strategies to find a middle ground between the two extremes~\cite{Grant2019initialization, GanzhornBarkoutsos19}, some adaptive approaches seek to construct the ansatz based on information gathered while running the algorithm. 
The earliest such algorithm was developed by Grimsley et al.~\cite{Grimsley2019ADAPT}, where, in contrast to conventional VQEs, the ansatz is grown dynamically instead of chosen a priori. This algorithm, Adaptive Derivative-Assembled Problem-Tailored Ansatz Variational Quantum Eigensolver (ADAPT-VQE), 
directly incorporates problem features into the ansatz, foregoing the maximum level of expressivity in favor of greater efficiency in the classical optimization, while at the same time retaining some degree of flexibility to avoid deep circuits.

The central idea of ADAPT-VQE is to add one operator at a time to the ansatz, with each new addition followed by classical optimization of the current ansatz as in conventional VQEs. Each new operator is selected from a predefined operator pool based on the gradient of the objective function, which is determined by the problem Hamiltonian together with the previously optimized state. This makes the ansatz inherently specific to the problem. 
It was shown that ADAPT-VQE results in shorter ans\"{a}tze with fewer variational parameters than UCCSD, except for certain strongly correlated molecules for which UCCSD fails to reach chemical accuracy.  Using a more hardware-friendly operator pool, qubit-ADAPT-VQE~\cite{qubitAdaptHoLun} requires even shallower circuits and fewer CNOT gates.
Moreover, by constructing a problem-specific ansatz, ADAPT-VQE is less prone to obstacles in the classical optimization compared to more hardware-efficient but problem-agnostic ans\"atze~\cite{GrimsleyRough}.

Despite the considerable reduction in circuit depth it offers compared to other VQEs, ADAPT-VQE still requires circuit depths on the order of tens of thousands to accurately estimate ground state energies of small molecules, which lies beyond the capacity of most quantum devices to this day~\cite{40yearsQC,Ladd2010}. This is due in part to the fact that each pool operator acts on only a small subset of qubits, leaving most qubits idle and unexploited at any given point in the construction of the state preparation circuit.
Moreover, the ansatz construction introduces additional overhead costs. Specifically, in each iteration of the algorithm, the gradient of the objective function needs to be measured for each operator in the pool. 
While this overhead is unlikely to be the bottleneck in near-term applications, it is desirable to reduce it in the long-run, when the system size of the computation task grows significantly.

In this work, we propose a variant of ADAPT-VQE, named Tiling Efficient Trial circuits with Rotations Implemented Simultaneously
(TETRIS)-ADAPT-VQE. We achieve the goal of further reducing the circuit depth by allowing unitary operators that act on disparate sets of qubits to be added to the ansatz simultaneously at each iteration. This results in a faster reduction of the objective function. 
We demonstrate this protocol using numerical simulations for a range of molecules with different geometries. 
We show that despite the much shallower quantum circuits, TETRIS-ADAPT can successfully produce results with the same accuracy as the original ADAPT-VQE.

The rest of the paper is organized as follows. In Sec.~\ref{sec:algorithm}, we review in detail the procedure of ADAPT-VQE, as it makes up the backbone of the algorithm we propose. We then present the new strategy, including a discussion of what operator pools to use and how to translate the sequence of unitaries in the ansatz to a circuit with quantum gates. Sec.~\ref{sec:results} contains the details and results of the numerical simulations that demonstrate the improvement TETRIS-ADAPT-VQE can provide. Compared to ADAPT-VQE, it further reduces the resources needed to reach a certain level of accuracy while keeping the advantages of ADAPT-VQE. Finally, we conclude in Sec.~\ref{sec:conclusion}.

\section{Algorithm details} \label{sec:algorithm}

Although our algorithm is quite general, in this work we focus on its application to problems in quantum computational chemistry. Computational chemistry is largely concerned with finding the ground state energy of the electronic part of molecular Hamiltonians:
\begin{equation}\label{hamilt}
    \hat{H}=\sum_{p,q} h_{pq}a_q^{\dagger}a_p+\frac{1}{2}\sum_{p,q,r,s} h_{pqrs}a_p^{\dagger}a_q^{\dagger}a_sa_r,
\end{equation}
where $h_{pq}$ and $h_{pqrs}$ are single- and two-electron integrals, and $a^\dagger$ and $a$ are the second-quantized fermionic creation and annihilation operators, respectively. 
Our algorithm takes the adaptive procedure as in ADAPT-VQE, with a novel strategy of selecting the operators that contribute to the VQE ansatz. Like ADAPT-VQE, the operator pools draw inspiration from the UCCSD ansatz, which is briefly reviewed in the next subsection. Readers familiar with the topic can jump to Sec.~\ref{subsec:algorithm-adapt}.

\subsection{Unitary Coupled Cluster}

The UCCSD ansatz has the form:
\begin{equation}
    \ket{\Psi_{\mathrm{UCCSD}}}=e^{\hat{T}_1+\hat{T}_2}\ket{\Psi_{\mathrm{ref}}},
\end{equation}
where
\begin{align}
    \hat{T}_1=\sum_{i,a} \hat{\tau}_i^a=\sum_{i,a} \tau_i^a(a_a^{\dagger}a_i-a_i^{\dagger}a_a), \label{singles} \\
    \hat{T}_2=\sum_{i,j,a,b} \hat{\tau}_{ij}^{ab}=\sum_{i,j,a,b} \tau_{ij}^{ab}(a_a^{\dagger}a_b^{\dagger}a_ia_j-a_j^{\dagger}a_i^{\dagger}a_ba_a). \label{doubles}
\end{align}
In practice the unitary is trotterized to first order:
\begin{equation}
     \ket{\Psi_{\mathrm{tUCCSD}}}=\prod_{c\in\{ia\}}e^{\hat{\tau}_c}\prod_{d\in\{ijab\}}e^{\hat{\tau}_d}\ket{\Psi_{\mathrm{ref}}},
\end{equation}
where $i,j$ ($a,b$) represent occupied (virtual) orbitals,
and the reference state $\ket{\Psi_{\mathrm{ref}}}$ is usually the Hartree-Fock state. From classical simulations, UCCSD is known to be a reliable ansatz, although it was recently shown~\cite{Grimsley2020} that the low-order trotterized form is not chemically well-defined as it fails to reach chemical accuracy for certain operator orderings. Furthermore, due to the large number of exponential factors, it is too expensive to prepare on quantum processors. Lastly, for strongly correlated systems, UCCSD often fails to reach chemical accuracy (error $<$ 1 kcal/mol) for the ground state energy~\cite{UCCrev}.

\subsection{ADAPT-VQE} \label{subsec:algorithm-adapt}

The ADAPT-VQE algorithm grows problem-tailored ans\"atze by appending one unitary at a time to the trial state~\cite{Grimsley2019ADAPT}. The user begins by defining an operator pool $\mathcal{P}=\{P_i\}$, a collection of antihermitian generators, such as the $\hat{\tau}_i^a$ and $\hat{\tau}_{ij}^{ab}$ in Eqs.~(\ref{singles}) and (\ref{doubles}), from which the ansatz is to be constructed. In addition, a reference state is chosen, usually the Hartree-Fock (HF) ground state. At each iteration, the expectation value of the commutator of the Hamiltonian and each pool operator is measured. This is equivalent to adding each candidate operator to the ansatz, setting its variational parameter equal to zero and computing the energy gradient with respect to it. The operator with the largest gradient norm is appended to the ansatz. 
For example, at the $(k+1)$-th layer, with the previously optimized state $\ket{\Psi^{(k)}}$, the energy gradient for the candidate operator $P_i$ with respect to its parameter $\theta_i$ is
\begin{align}
    \frac{\partial E}{\partial \theta_i}\bigg|_{\theta_i=0} &= \left[\frac{\partial}{\partial \theta_i} \bra{\Psi^{(k)}}e^{-\theta_iP_i}He^{\theta_iP_i}\ket{\Psi^{(k)}}\right]
    \bigg|_{\theta_i=0} \nonumber \\
    &= \bra{\Psi^{(k)}}[H,P_i]\ket{\Psi^{(k)}},
\end{align}
using the antihermiticity $P_i^\dagger=-P_i$.
The algorithm proceeds as follows:
\begin{enumerate}
    \item On the classical device, compute one- and two-body integrals and transform the fermionic Hamiltonian to a qubit Hamiltonian using a suitable mapping, e.g., Jordan-Wigner.
    \item On the quantum device, prepare the current ansatz and measure the energy gradient $\frac{\partial E}{\partial \theta_i}\rvert_{\theta_i=0}$ for every candidate pool operator $P_i$ with respect to its variational parameter $\theta_i$.
    \item If the pool gradient norm is smaller than a predetermined threshold, ADAPT-VQE has converged, and the algorithm terminates. 
    \item Add the operator with the largest gradient norm from step 2 to the ansatz, with its variational parameter set to zero.
    \item Perform a VQE subroutine to update \textit{all} parameters in the current ansatz.
    \item Repeat steps 2 -- 5 until convergence.
\end{enumerate}

\subsection{TETRIS-ADAPT-VQE} \label{subsec:algorithm-tetris}

\begin{figure}
  \includegraphics[width=1\linewidth]{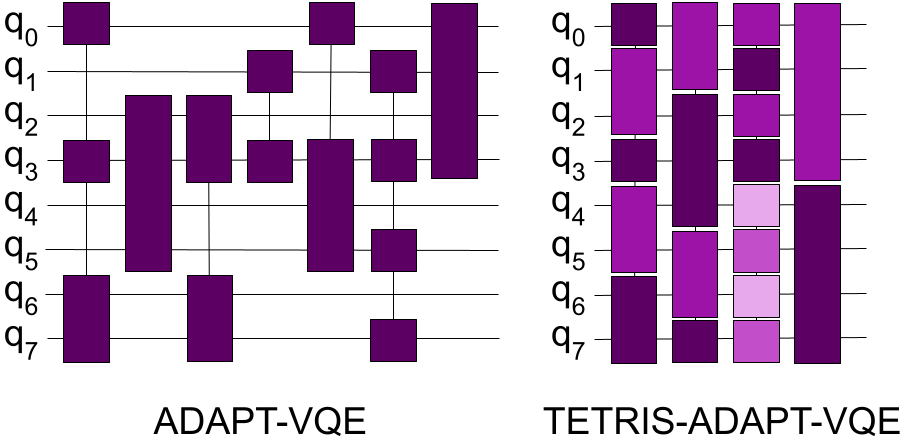}
  \caption{Schematic diagram of the ansatz preparation circuits produced by ADAPT-VQE (left) and TETRIS-ADAPT-VQE (right). TETRIS-ADAPT-VQE grows ans\"atze prepared by shallower and denser circuits compared to ADAPT-VQE, by adding multiple operators at each iteration. The dark violet rectangles represent gates implementing the largest-gradient operator, whereas lighter tones correspond to operators beyond the highest-gradient one, added in TETRIS-ADAPT-VQE.}
  \label{fig:regvstetris}
\end{figure}

In TETRIS-ADAPT-VQE, we add not only the operator associated with the largest gradient, but also the operator associated with the next largest gradient norm that is supported on qubits \textit{different} from the support of the first added operator. We continue this process of adding operators with successively smaller (or equal) gradients and with support that is disjoint from that of all previously added operators, until no further operators satisfying these criteria are found. Note that this can be done without measuring any additional gradients compared to ordinary ADAPT-VQE since the gradient of each pool operator need only be measured once per iteration of the algorithm as usual. 
This modification of the algorithm amounts to replacing the original step 4 from above with the following subroutine for adding operators to the ansatz:
\begin{enumerate}
    \item [4'.]
        \begin{enumerate}
            \item Sort pool operators according to the norms of their gradients, in descending order.
            \item Identify the operator with the highest gradient norm acting on qubits not acted on by any previously added operator in the current ADAPT iteration. Add the operator to the ansatz, with its variational parameter set to zero.
            \item If operators that act non-trivially on all qubits have been added to the ansatz in the current ADAPT iteration, proceed to step 5.
            \item If the list of operators with nonzero gradients and disjoint support with operators already added in the current ADAPT iteration has been exhausted, proceed to step 5.
            \item Go to sub-step (b).
        \end{enumerate}
\end{enumerate}
We illustrate this strategy in Fig.~\ref{fig:regvstetris}, showing the locations of the gates implementing the operators chosen by the original ADAPT-VQE and TETRIS-ADAPT-VQE. Note that the highest-gradient operators coincide for the two versions of the algorithm in the first layer, but differ in the subsequent layers, as a consequence of the diverging ans\"atze.

\subsection{Operator Pools}

The original ADAPT-VQE paper~\cite{Grimsley2019ADAPT} compared ADAPT ans\"atze to the commonly used, fixed UCCSD ansatz, and as such it used pools of antihermitian sums of single and double fermionic excitation-deexcitation operators of the form $\hat{T}_{ij}=a_i^{\dagger}a_j-a_j^{\dagger}a_i$ and $\hat{T}_{ijkl}=a_i^{\dagger}a_j^{\dagger}a_ka_l-a_l^{\dagger}a_k^{\dagger}a_ja_i$. In the Jordan-Wigner mapping, which we use in the remainder of this paper, $\hat{T}_{ij}$ and $\hat{T}_{ijkl}$, for spin-orbital (and qubit) indices $i<j<k<l$, can be written in terms of sums of Pauli strings as
\begin{align}
    \hat{T}_{ij}&=\frac{i}{2}(X_iY_j-Y_iX_j)\prod_{p=i+1}^{j-1}Z_p, \label{single_excitation} \\
    \hat{T}_{ijkl}&=\frac{i}{8}(X_iY_jX_kX_l+Y_iX_jX_kX_l+Y_iY_jY_kX_l \nonumber \\
    &+Y_iY_jX_kY_l-X_iX_jY_kX_l-X_iX_jX_kY_l \nonumber \\
    &-Y_iX_jY_kY_l-X_iY_jY_kY_l)\prod_{p=i+1}^{j-1}Z_p\prod_{r=k+1}^{l-1}Z_r, \label{double_excitation}
\end{align}
where the Pauli strings are tensor products of Pauli matrices on the corresponding qubits, and identities are implied on qubits with omitted indices. Although chemically sound, the large number of terms and linearly increasing number of qubits involved in these operators translate into deep circuits with large numbers of two-qubit gates (see the following subsection), and make fermionic operators challenging for NISQ simulation. 

In an effort to construct even more hardware-efficient ans\"atze, Tang et al.~\cite{qubitAdaptHoLun} showed that decomposing the Jordan-Wigner mapped fermionic operators into individual Pauli strings and using those as pool operators results in shallower circuits, an approach dubbed qubit-ADAPT-VQE. In the same work, it was found that omitting the trailing Pauli-Zs from the Pauli strings did not affect convergence. At the expense of additional variational parameters, qubit-ADAPT-VQE yielded ans\"atze with up to an order of magnitude fewer CNOT gates for the systems studied.

In a similar spirit, Yordanov et al.~\cite{Yordanov2021QEB} introduced the Qubit-Excitation-Based (QEB)-ADAPT-VQE, which uses single and double qubit excitations, obtained by replacing the fermionic creation and annihilation operators by the corresponding qubit operators $Q_i^\dagger=(X_i-iY_i)/2$ and $Q_i=(X_i+iY_i)/2$ satisfying
\begin{align}
    \{Q_i,Q_i^\dagger\}=1,\; [Q_i, Q_j^\dagger]=0\;\hbox{for} \;i\neq j,\label{qubitcommute}\\
    [Q_i,Q_j]=[Q_i^\dagger,Q_j^\dagger]=0. 
\end{align}
In light of Eq.~(\ref{qubitcommute}), in the Jordan-Wigner mapping, this is equivalent to omitting the Pauli-Z strings in Eqs.~(\ref{single_excitation}) and (\ref{double_excitation}). Using the optimized circuits in Ref.~\cite{YordanovOptimizedCircuits} to implement the qubit excitation (QE) evolutions resulted in ans\"atze with CNOT counts similar or lower than those of qubit-ADAPT-VQE and numbers of variational parameters comparable to those of fermionic ADAPT-VQE.

In this work we use the qubit and QE pools because their operators act non-trivially on fewer qubits compared to those of the fermionic pool. In general, more operators can therefore be added at each TETRIS-ADAPT-VQE iteration. 
The qubit pool has generators of the form 
\begin{equation}
    iX_iY_j, \, iX_iX_jX_kY_l, \, iY_iY_jY_kX_l,
\end{equation}
and we include all combinations of qubit indices such that their sum is an even number.
In the QE pool, the qubit-excitation generators for $\hat{\overline{T}}_{ij}=Q_i^{\dagger}Q_j-Q_j^{\dagger}Q_i$ and $\hat{\overline{T}}_{ijkl}=Q_i^{\dagger}Q_j^{\dagger}Q_kQ_l-Q_l^{\dagger}Q_k^{\dagger}Q_jQ_i$ in the Jordan-Wigner mapping have the form: 
\begin{align} 
    \hat{\overline{T}}_{ij}&=\frac{i}{2}(X_iY_j-Y_iX_j), \label{single_q_excitation} \\
    \hat{\overline{T}}_{ijkl}&=\frac{i}{8}(X_iY_jX_kX_l+Y_iX_jX_kX_l \nonumber \\
    &+Y_iY_jY_kX_l+Y_iY_jX_kY_l-X_iX_jY_kX_l \nonumber \\
    &-X_iX_jX_kY_l-Y_iX_jY_kY_l-X_iY_jY_kY_l), \label{double_q_excitation}
\end{align}
where we include all combinations of indices such that no spin flips are allowed. For both pools it is implied that $i\neq j \neq k \neq l$, and many index permutations in the operators above differ only by an overall sign and need not be included in practice.

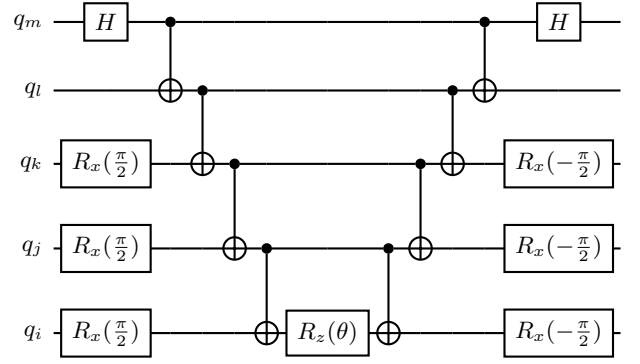
\begin{figure}
\begin{tikzpicture}
\node[scale=1] {
\begin{quantikz}[column sep=0.1cm]
\lstick{$q_m$}&\gate{H}&\ctrl{1}&\qw&\qw&\qw&\qw&\qw&\qw&\qw&\ctrl{1}&\gate{H}&\qw\\
\lstick{$q_l$}&\qw&\targ{}&\ctrl{1}&\qw&\qw&\qw&\qw&\qw&\ctrl{1}&\targ{}&\qw&\qw\\
\lstick{$q_k$}&\gate{R_x(\frac{\pi}{2})}&\qw&\targ{}&\ctrl{1}&\qw&\qw&\qw&\ctrl{1}&\targ{}&\qw&\gate{R_x(-\frac{\pi}{2})}&\qw\\
\lstick{$q_j$}&\gate{R_x(\frac{\pi}{2})}&\qw&\qw&\targ{}&\ctrl{1}&\qw&\ctrl{1}&\targ{}&\qw&\qw&\gate{R_x(-\frac{\pi}{2})}&\qw\\
\lstick{$q_i$}&\gate{R_x(\frac{\pi}{2})}&\qw&\qw&\qw&\targ{}&\gate{R_z(\theta)}&\targ{}&\qw&\qw&\qw&\gate{R_x(-\frac{\pi}{2})}&\qw\\
\end{quantikz}
};
\end{tikzpicture}

\caption{\label{fig:expmap}Example circuit of the exponential map implementing $e^{-i\frac{\theta}{2} Y_iY_jY_kZ_lX_m}$.}
\end{figure}

\subsection{Quantum Circuits} \label{subsec:algorithm-circ}

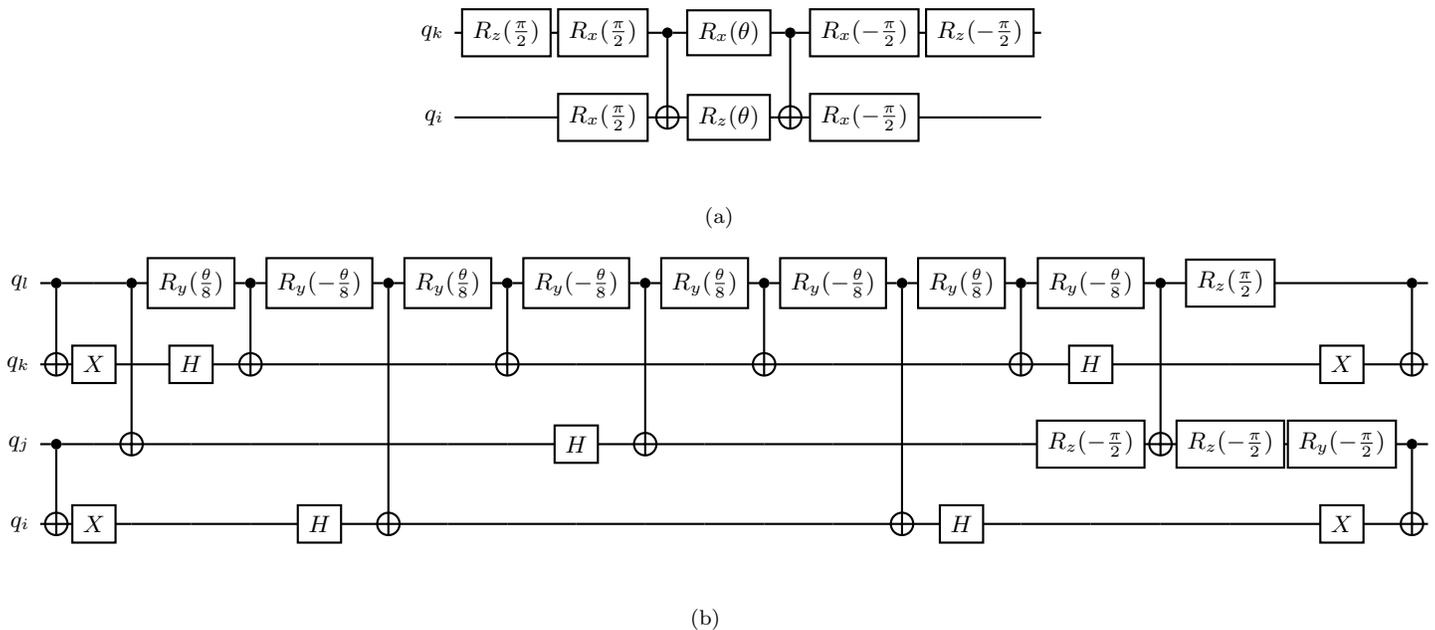
\begin{figure*}

\subfigure[]{
\begin{tikzpicture}
\node[scale=1] {
\begin{quantikz}[column sep=0.1cm]
\lstick{$q_k$}&\gate{R_z(\frac{\pi}{2})}&\gate{R_x(\frac{\pi}{2})}&\ctrl{1}&\gate{R_x(\theta)}&\ctrl{1}&\gate{R_x(-\frac{\pi}{2})}&\gate{R_z(-\frac{\pi}{2})}&\qw\\
\lstick{$q_i$}&\qw&\gate{R_x(\frac{\pi}{2})}&\targ{}&\gate{R_z(\theta)}&\targ{}&\gate{R_x(-\frac{\pi}{2})}&\qw&\qw\\
\end{quantikz}
};
\end{tikzpicture}
}

\begin{adjustwidth}{-1cm}{}
\subfigure[]{
\begin{tikzpicture}
\node[scale=1] {
\begin{quantikz}[column sep=0.05cm]
\lstick{$q_l$}&\ctrl{1}&\qw&\ctrl{2}&\gate{R_y(\frac{\theta}{8})}&\ctrl{1}&\gate{R_y(-\frac{\theta}{8})}&\ctrl{3}&\gate{R_y(\frac{\theta}{8})}&\ctrl{1}&\gate{R_y(-\frac{\theta}{8})}&\ctrl{2}&\gate{R_y(\frac{\theta}{8})}&\ctrl{1}&\gate{R_y(-\frac{\theta}{8})}&\ctrl{3}&\gate{R_y(\frac{\theta}{8})}&\ctrl{1}&\gate{R_y(-\frac{\theta}{8})}&\ctrl{2}&\gate{R_z(\frac{\pi}{2})}&\qw&\ctrl{1}&\qw\\
\lstick{$q_k$}&\targ{}&\gate{X}&\qw&\gate{H}&\targ{}&\qw&\qw&\qw&\targ{}&\qw&\qw&\qw&\targ{}&\qw&\qw&\qw&\targ{}&\gate{H}&\qw&\qw&\gate{X}&\targ{}&\qw\\
\lstick{$q_j$}&\ctrl{1}&\qw&\targ{}&\qw&\qw&\qw&\qw&\qw&\qw&\gate{H}&\targ{}&\qw&\qw&\qw&\qw&\qw&\qw&\gate{R_z(-\frac{\pi}{2})}&\targ{}&\gate{R_z(-\frac{\pi}{2})}&\gate{R_y(-\frac{\pi}{2})}&\ctrl{1}&\qw\\
\lstick{$q_i$}&\targ{}&\gate{X}&\qw&\qw&\qw&\gate{H}&\targ{}&\qw&\qw&\qw&\qw&\qw&\qw&\qw&\targ{}&\gate{H}&\qw&\qw&\qw&\qw&\gate{X}&\targ{}&\qw\\
\end{quantikz}
};
\end{tikzpicture}
}
\end{adjustwidth}

\caption{\label{fig:QEcirc} Example circuits performing (a) a single qubit excitation, $e^{i\frac{\theta}{2}(X_iY_k-Y_iX_k)}$, and (b) a double qubit excitation evolution, $e^{\theta\hat{\overline{T}}_{ijkl}}$, as in Eq.~\eqref{double_q_excitation}.}
\end{figure*}

To obtain the CNOT gate counts and circuit depths we compile circuits effecting the operator evolutions in Qiskit~\cite{Qiskit}. For the qubit pool, which consists of individual Pauli strings, the exponentiation is performed with the usual exponential map circuit. For example, suppose that the operator $Z_a\otimes Z_b \otimes Z_c$ is to be exponentiated. Then a ``CNOT staircase" computes the parities of the three qubits in the computational basis, an $R_z(\theta)$ gate applies the phase shift, and an inverse CNOT staircase uncomputes the parity. If the tensor product in the exponent contains $X$ or $Y$ Paulis, then those are rotated to the $Z$ basis by $H$ and $R_x(\frac{\pi}{2})$ gates respectively. Fig.~\ref{fig:expmap} shows the circuit performing $e^{-i \frac{\theta}{2} Y_iY_jY_kZ_lX_m}$.

For the QE pool, we use the optimized circuits introduced in Refs. \cite{YordanovOptimizedCircuits, Yordanov2021QEB}, derived by sequentially decomposing into opposite half-rotations the multi-qubit-controlled gates in exchange-interaction-type circuits, noting that single and double qubit excitation operators $\hat{\overline{T}}_{ij}$ and $\hat{\overline{T}}_{ijkl}$ continuously exchange states: $\ket{1_i0_j}$ with $\ket{0_i1_j}$ and $\ket{1_i1_j0_k0_l}$ with $\ket{0_i0_j1_k1_l}$. We show two example circuits in Fig.~\ref{fig:QEcirc} taken from Ref.~\cite{Yordanov2021QEB}. Throughout this work, we assume all-to-all qubit connectivity, so CNOT gates between non-neighboring qubits can be performed.

\section{Results} \label{sec:results}

We numerically simulate the performances of TETRIS-ADAPT-VQE and the original ADAPT-VQE for the tasks of finding the ground state energy and preparing the ground state of the following molecules: \ch{H4}, \ch{LiH}, \ch{H6}, and \ch{BeH2} at varying bond lengths, all in the linear configuration, and using the STO-3G basis set. We will show that TETRIS-ADAPT-VQE provides additional improvement on both circuit depth and measurement overhead, without sacrificing the merits of the original ADAPT-VQE.
The results in this section were produced using python code written by the authors. The classical optimization scheme used throughout this work is the BFGS method as implemented in SciPy,~\cite{2020SciPy-NMeth} with a gradient norm tolerance of $10^{-10}$. The ADAPT convergence criterion was a pool gradient norm threshold of $10^{-7}$. All energies are in units of Hartrees.

\subsection{Circuit depth reduction}

The main motivation of TETRIS-ADAPT-VQE is circuit depth reduction. Since ADAPT-VQE only adds the largest gradient operator at each iteration, one may speculate that adding operators with smaller gradients alongside it would result in sub-optimal ans\"atze with a greater total number of parameters and CNOT gates. On the other hand, the magnitude of the operator gradient, which is measured about the point in parameter space where the VQE converged in the previous iteration, contains only local information about the ability of a candidate operator to reduce the value of the cost function. Put another way, the operator with the highest gradient is not guaranteed to be the operator that causes the greatest energy reduction. In Ref. \cite{Yordanov2021QEB} the authors explored the possibility of separately appending the $n$ largest-gradient operators to the ansatz at each ADAPT iteration, performing VQE on all $n$ trial ans\"atze, and updating the main ansatz with only the operator that resulted in the lowest energy. 
Moreover, adding operators beyond the highest-gradient one may enable further adjustments to the ansatz that would otherwise appear in later layers. In fact, we show below that TETRIS-ADAPT-VQE achieves circuit depth reductions with numbers of variational parameters and CNOT gates similar to those required by the original ADAPT-VQE.

\begin{figure*}
\centering
\includegraphics[width=16cm, height=16cm, keepaspectratio]{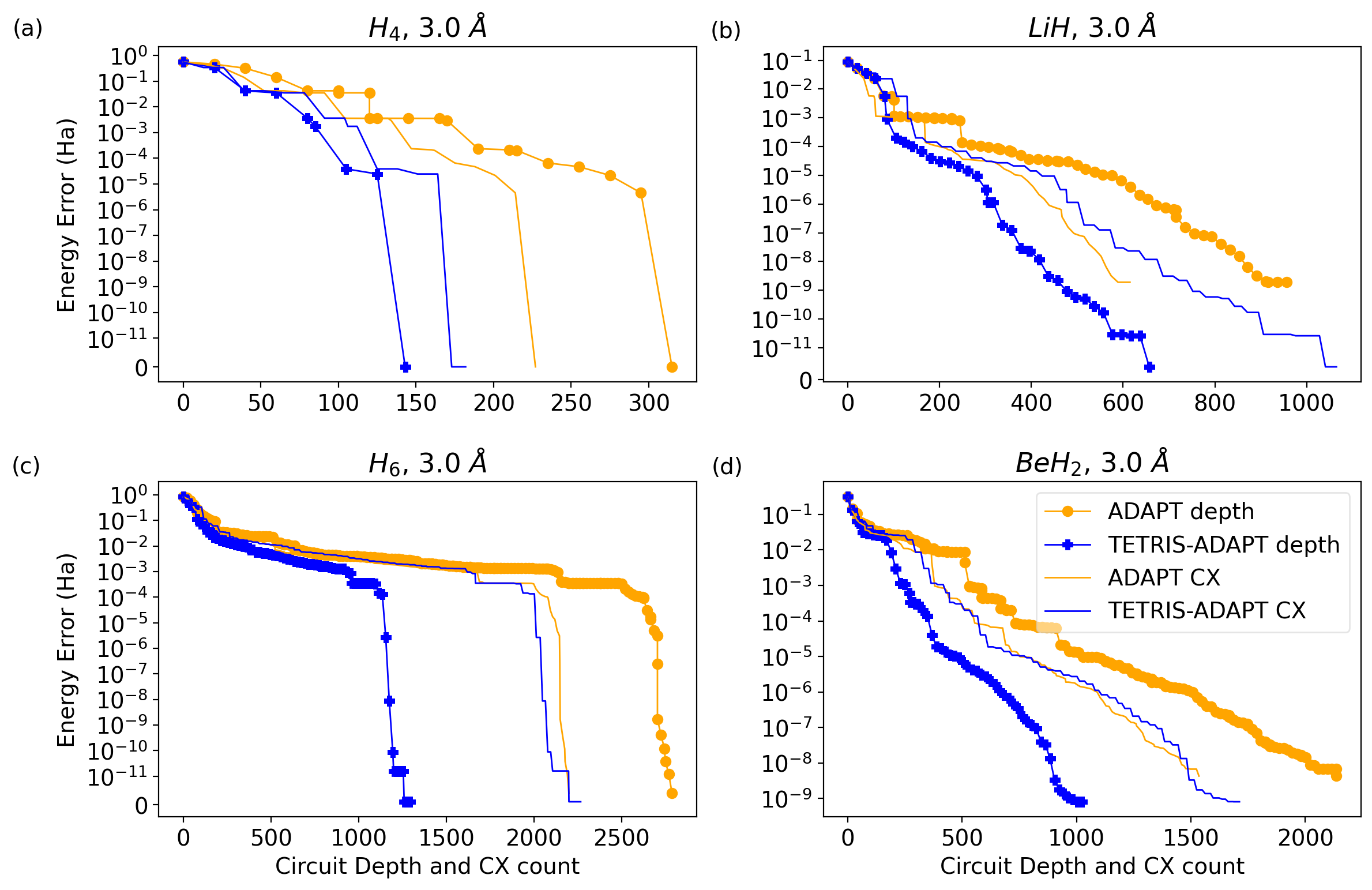}
\caption{Resources required by ADAPT-VQE (orange lines) and TETRIS-ADAPT-VQE (blue lines). Circuit depths and CNOT counts as functions of energy error for linear (a) \ch{H4}, (b) \ch{LiH}, (c) \ch{H6}, and (d) \ch{BeH2} with internuclear distances of 3.0 \AA~and the QE pool.}
\label{fig:circuit_depth}
\end{figure*}

In this subsection, we compare ADAPT-VQE and TETRIS-ADAPT-VQE in terms of CNOT counts and circuit depths. Fig.~\ref{fig:circuit_depth} shows the results for linear \ch{H4}, \ch{LiH}, \ch{H6}, and \ch{BeH2} at an internuclear distance of 3.0 \AA, using the QE pool. CNOT counts and circuit depths are obtained from circuits constructed as described in Sec.~\ref{subsec:algorithm-circ} using built-in Qiskit functions, and Qiskit optimization level 1 (back-to-back gate cancellation). We note that for a fair comparison, when transpiling ADAPT-VQE circuits, unitaries added in consecutive layers but acting on disparate sets of qubits are performed concurrently.

It is evident that TETRIS-ADAPT-VQE leads to ans\"atze prepared by much shallower circuits, while requiring roughly the same number of CNOT gates as ADAPT-VQE. Furthermore, the larger the system size, the greater the advantage of TETRIS-ADAPT-VQE as shown in Table \ref{table:CircuitDepthSummary}, because more operators fit in the same ADAPT layer, in general.

\begin{table}
\begin{tabular}{c|c|c|c}

    \hline Molecule& Qubits & Pool   & Avg. Circuit Depth Ratio    \\\hline\hline
    \ch{H4}&8      & qubit   &  1.64 \\
    \ch{H4}&8      & QE          & 1.58 \\
    \ch{LiH}&12      & qubit     &  2.08 \\
    \ch{LiH}&12      & QE   &  1.76    \\
    \ch{H6}&12       & qubit     &  2.32 \\
    \ch{H6}&12       & QE      &  2.25  \\
    \ch{BeH2}&14     & qubit     &  2.73 \\
    \ch{BeH2}&14     & QE    &  2.56 \\\hline
\end{tabular}
\caption{\label{table:CircuitDepthSummary}Summary of circuit depth results for the four molecules and the two pools we study. The ADAPT-VQE to TETRIS-ADAPT-VQE  average (over different bond lengths) circuit depth ratio calculated at ADAPT-VQE convergence is given in the fourth column. Geometries for which either version of the algorithm failed to reach the ground state were excluded. The number of qubits required to simulate each molecule is given in column 2. It is evident that the ratio grows with the system size. The ratio also seems to be consistently higher for the qubit pool.} 
\end{table}

\begin{figure*}
\centering
\includegraphics[width=16cm, height=8cm, keepaspectratio]{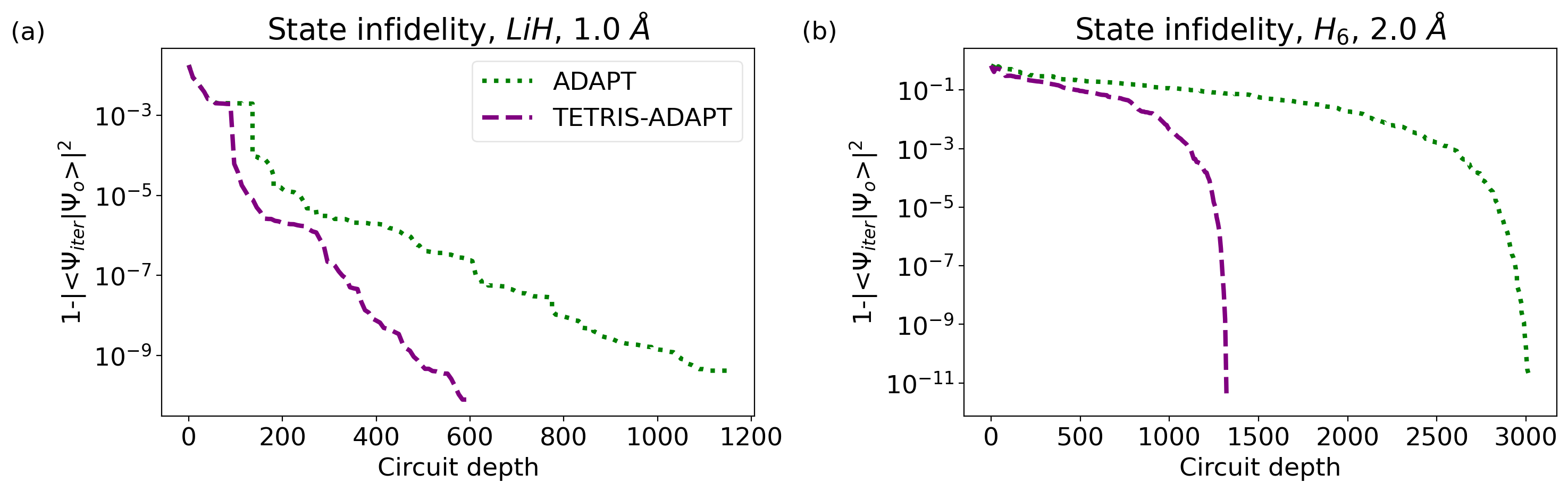}
\caption{Performances of ADAPT-VQE (green dotted line) and TETRIS-ADAPT-VQE (purple dashed line) in terms of state infidelity as a function of circuit depth for (a) \ch{LiH}, with 1.0 \AA\ bond length and (b) linear \ch{H6} with 2.0 \AA\ bond length and the qubit pool.}
\label{fig:StateInfidelity}
\end{figure*}

Beyond ground-state energy estimation, ADAPT-VQE can be used for ground-state preparation. The faster energy descent of TETRIS-ADAPT does not come at an expense of ansatz quality, as the two versions of the algorithm yield equally faithful ans\"atze. Fig.~\ref{fig:StateInfidelity} shows that the TETRIS version of the algorithm approximates the true ground state significantly faster than the original implementation.

\subsection{Measurement overhead reduction}

In the NISQ era, the bottleneck for VQEs is the circuit depth and number of gates required for state preparation, rather than the number of state preparations and measurements. Although ADAPT-VQE yields more compact ans\"atze compared to static ones such as the tUCCSD for the same energy accuracy, it does so at the additional cost of repeatedly measuring the pool operator gradient expectation values. This step can be performed in parallel, assuming the availability of many quantum processors. Still, we would like to reduce the quantum resource requirements associated with the gradient measurement step. One way to go about this is to reduce the number of operators in the chosen operator pool by removing redundancies and exploiting known symmetries of the system under study. See Refs~\cite{qubitAdaptHoLun, VladSymmetry} for discussions on operator pool redundancy, minimal complete pools, and symmetry considerations.
Another approach to minimize the measurement overhead has been proposed by Yang and coworkers~\cite{yang2021RDM}, where they use an approximate reconstruction of the three-electron reduced density matrix to completely remove the measurment overhead, albeit at a slight increase in ansatz length. TETRIS-ADAPT-VQE adds multiple operators per ADAPT layer while keeping the number of operators roughly constant. Thus, it reduces the total number of layers and by extension the number of times the operator pool gradients are measured, by a factor that grows roughly linearly with the system size. Assuming that the ansatz is dominated by double excitation-like operators, which act on 4 qubits each for the qubit and QE pools in the JW mapping, this factor is expected to be roughly $\lfloor\frac{Q}{4}\rfloor$ where $Q$ is the number of qubits. In Table~\ref{table:measurements}, we list the measurement overhead reduction observed in the simulation of the four molecules.

\begin{table}[!ht]
    \centering
    \begin{tabular}{c|c|c|c}\hline
        Molecule & Qubits & Percentage reduction & Ratio \\\hline\hline
        \ch{H4} & 8 & 53\% & 2.1 \\
        \ch{LiH} & 12 & 59\%  & 2.5 \\
        \ch{H6} & 12 & 66\%  & 3 \\
        \ch{BeH2} & 14 & 70\%  & 3.3 \\\hline
        \end{tabular}
\caption{Percentage reduction of the number of operator gradient measurements, of TETRIS-ADAPT-VQE relative to ADAPT-VQE calculated at ADAPT-VQE convergence, and the corresponding ratio. No operator grouping for simultaneous measurement has been assumed in these results. Results are averaged over all molecular geometries simulated, excluding those for which either version of the algorithm failed to reach the ground state.}
\label{table:measurements}    
\end{table}

\subsection{Enhanced ansatz expressivity}

Simply by adding multiple operators acting on qubits different from the support of the highest-gradient operator, the TETRIS strategy lends ADAPT-VQE the ability to explore a larger part of the Hilbert space, at least at earlier stages of the algorithm. The ansatz can therefore overlap with some components in the true ground state that are otherwise inaccessible to a shallow ansatz.

We now use a concrete example to illustrate this. We consider the ans\"atze grown for \ch{H4} at $3.0$ \AA\ using the qubit pool, with the HF reference state, $\ket{\Psi_{\mathrm{ref}}}=\ket{11110000}$. The optimized qubit-ADAPT ans\"atze after one and two iterations of the algorithm are, respectively, 
\begin{align*}
    & \ket{\phi_1}=e^{0.5652iX_2X_3X_6Y_7}\ket{\Psi_{\mathrm{ref}}}\\&= 0.8445\ket{11110000}-0.5356\ket{11000011},
\end{align*}
and 
\begin{align*}
    &\ket{\phi_2}=e^{-1.5708iX_0X_3X_5Y_6}e^{1.5708iX_2X_3X_6Y_7}\ket{\Psi_{\mathrm{ref}}}\\&=\ket{01010101}.
\end{align*}
In this specific example, qubit-ADAPT arrives in the all spin-down state, an excited state of the system, and the ansatz growth stops, as the pool gradient vanishes at eigenstates of the Hamiltonian. 
For the same system, the corresponding TETRIS-qubit-ADAPT-VQE ansatz after just one iteration is
\begin{align*}
    &\ket{\psi_1}=e^{0.6757iX_0X_1X_4Y_5+0.6748iX_2X_3X_6Y_7}\ket{\Psi_{\mathrm{ref}}}
    \\&=0.6092\ket{11110000}-0.4884\ket{00111100}\\&-0.4875\ket{11000011}+0.3908\ket{00001111},
\end{align*}
where adding two double-excitation-like qubit operators side-by-side generates a quadruple excitation subterm, i.e., the $\ket{00001111}$ component. The quadruply excited Slater determinant turns out to be the most dominant term beyond the reference determinant in the ground state:
\begin{align*}
    &\ket{\Psi_{0}}=0.5182\ket{11110000}+0.4429\ket{00001111}\\&-0.3658\ket{11001100}-0.3394\ket{00110011}+\dots,
\end{align*}
where $\ket{\Psi_0}$ is the ground state, and the computational basis states are ordered according to the magnitude of their coefficients. In this case, the TETRIS strategy enables the ans\"atze to include this dominant term within just the first layer, whereas the original qubit-ADAPT ansatz with one layer is orthogonal to $\ket{00001111}$. TETRIS-qubit-ADAPT-VQE converges to the exact ground state within 13 iterations. 

\begin{figure}
  \includegraphics[width=1\linewidth]{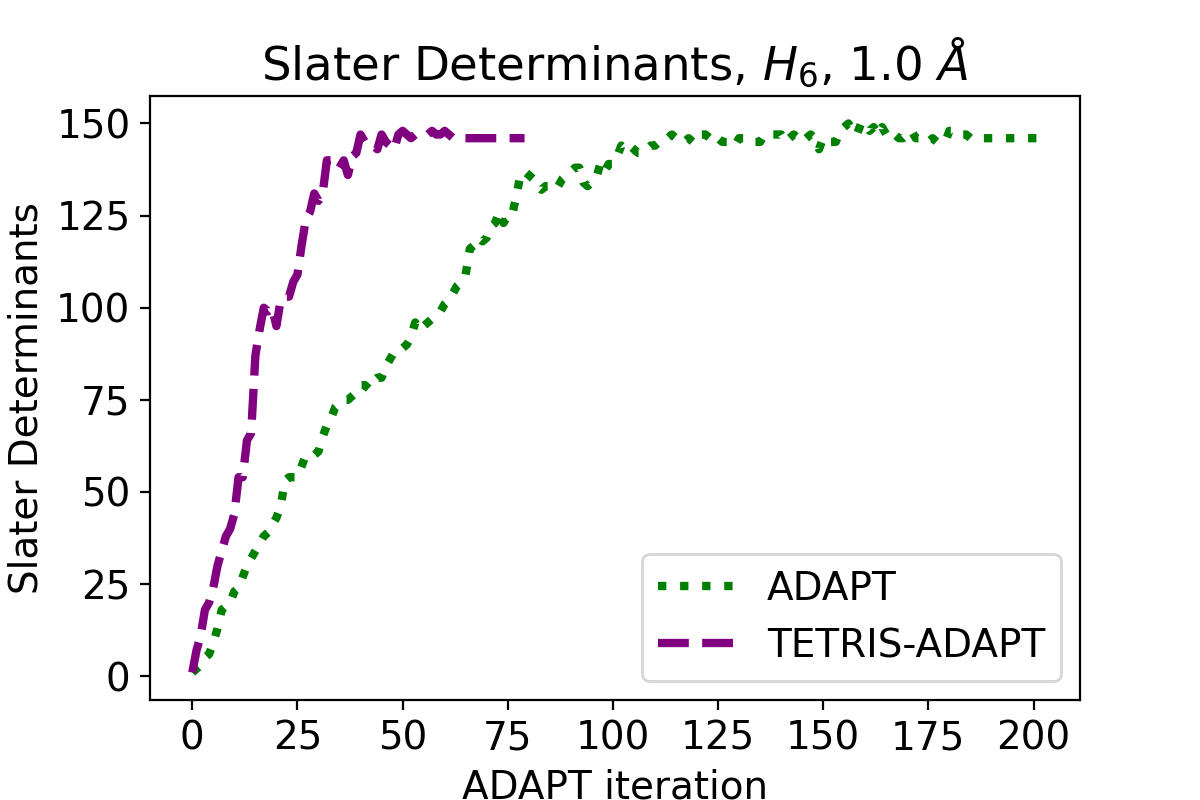}
  \caption{Number of Slater determinants in the ansatz (equivalently, computational basis states in the state of the quantum processor) with coefficients of absolute value greater or equal to $10^{-3}$, as a function of ADAPT iteration for linear \ch{H6}, bond length of 1.0 \AA\, and the QE pool.}
  \label{fig:slaterdeterminants}
\end{figure}

We argue that the increase in expressivity here is still \textit{problem-aware}, as the operators are added in the order of the magnitude of their gradients. This is in contrast to the problem-agnostic ansatz where entangling gates are blindly added to enlarge the accessible Hilbert space. Fig. \ref{fig:slaterdeterminants} shows the number of Slater determinants in the ansatz as a function of ADAPT layer for \ch{H6} at $1.0$ \AA\ and both versions of the algorithm. TETRIS-ADAPT-VQE explores the Hilbert space faster, but just the right amount, as both curves plateau at the same number. We present numerical evidence in Sec.~\ref{subsec:results-opt} to demonstrate that the TETRIS strategy does not lose the remarkable optimizability of ADAPT-VQE.

\subsection{Optimizability} \label{subsec:results-opt}

The optimizability of ADAPT-VQE and the parameter landscape it generates were recently investigated in Ref. \cite{GrimsleyRough} by searching for local minima via random parameter initializations and subsequent optimizations at each ADAPT iteration. By virtue of adding a large gradient operator to the ansatz at each layer, ADAPT-VQE generates a steep parameter landscape near the energy minimum found in the previous iteration. Furthermore, using the parameters found in the previous VQE step as the starting point in the new iteration ensures that ADAPT-VQE focuses on one (possibly not global) energy minimum, thus ``burrowing" through the parameter landscape. 
Because the extra operators TETRIS-ADAPT-VQE adds to the ansatz at each layer do not interfere with the largest-gradient one, they do not disrupt the steep parameter landscape. Instead, they provide additional directions in which the optimizer can march, thus increasing the chance of escaping a high-energy trap. For these reasons, we expect TETRIS-ADAPT-VQE to inherit the favourable features of the original ADAPT-VQE.

We employ the same technique to show that TETRIS-ADAPT-VQE possesses the same desirable features as ADAPT-VQE. Choosing the molecules \ch{LiH} (at 3.0 \AA) and linear \ch{H6} (at 1.0 \AA) as examples, we simulate ADAPT-VQE and TETRIS-ADAPT-VQE with the QE pool and different parameter initialization schemes. At each ADAPT iteration, 300 random parameter initializations are drawn from a uniform distribution from $-\pi$ to $+\pi$ as the initial parameters in the ansatz, which lead to local energy minima found in subsequent optimization using the BFGS algorithm. These local minima are compared to the minimum energies found using two other initializations: setting the newly added parameters to zero while recycling the previously found optimal parameters, which we refer to as ``warm start", and setting all the parameters in the ansatz to zero, referred to as ``cold start".

\begin{figure*}
\centering
\includegraphics[width=16cm, height=16cm, keepaspectratio]{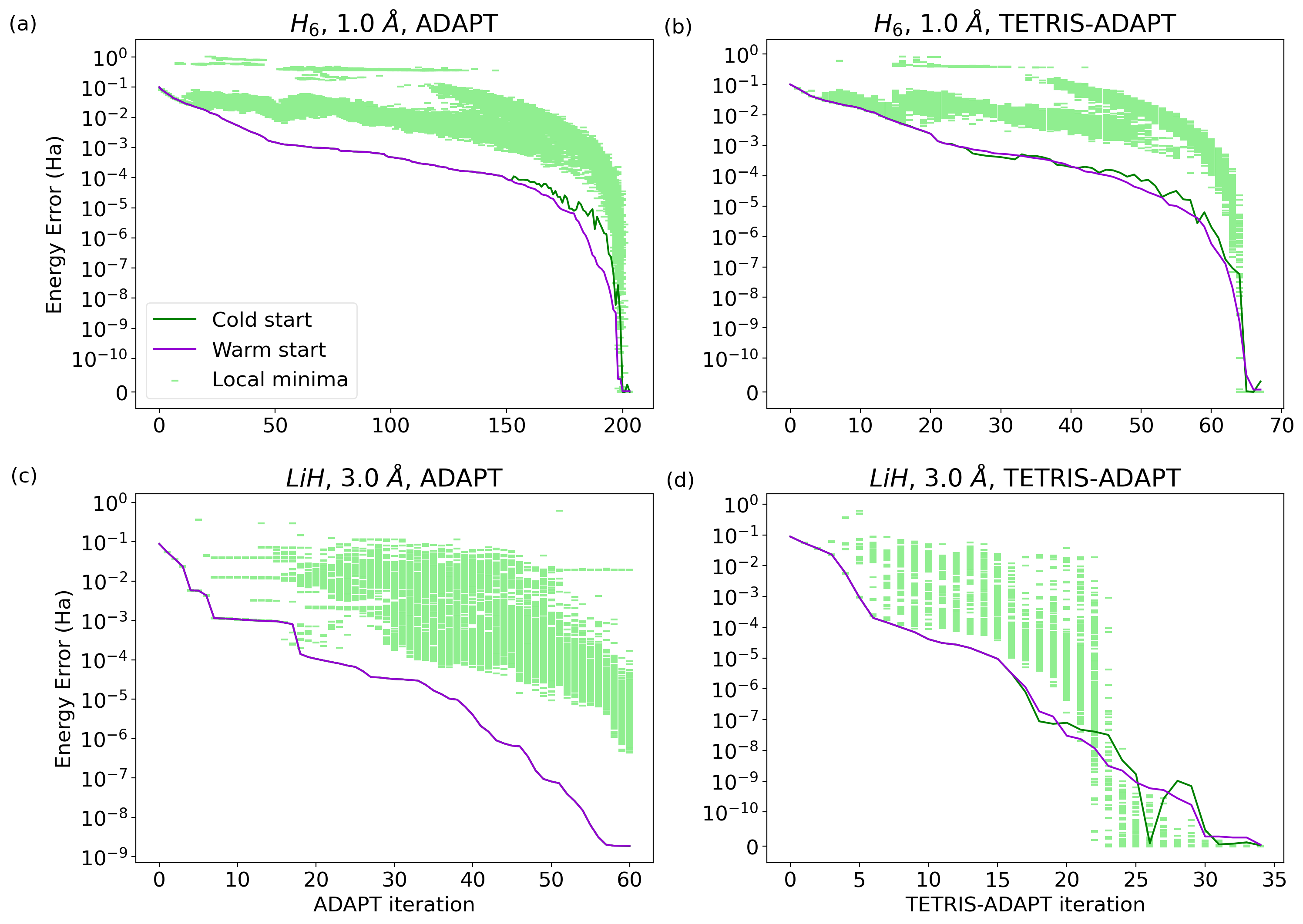}

\caption{Convergence curves for (a),(b) \ch{H6} at $1.0$ \AA\ and (c),(d) \ch{LiH} at $3.0$ \AA\ for (a),(c) ADAPT-VQE and (b),(d) TETRIS-ADAPT-VQE using the QE pool. Energy minima found by random parameter initialization and subsequent optimization at each ADAPT iteration are shown in light green. The warm start initialization in which parameters are initialized in their optimal values from the previous ADAPT iteration is shown (purple line). The all-zero parameter initialization (cold start) is also shown (solid green line). } 
\label{fig:localminima}
\end{figure*}

From Fig.~\ref{fig:localminima}, it is evident that as the dimension of the parameter space increases, it becomes increasingly improbable for random initializations to find the minima reached by both versions of ADAPT-VQE and warm-starting. We also note that although cold-starting is remarkably good at finding the ADAPT-VQE minimum, it also requires a greater number of function evaluations.

This demonstrates the efficacy of the usual initialization (i.e. warm start) in ADAPT-VQE, regardless of whether or not the TETRIS strategy is used. With this practical initialization scheme, the classical optimization in TETRIS-ADAPT-VQE produces a result equally as good as that of ADAPT-VQE with a similar number of variational parameters, even though at a given layer the dimension of the parameter space increases by more than one.

\section{Conclusions} \label{sec:conclusion}

Quantum simulation algorithms designed for NISQ devices should account for limited coherence times and noisy gates. In prior work, ADAPT-VQE, which iteratively grows problem-tailored ans\"atze based on local energy gradient information, was shown to yield accurate trial wave functions requiring fewer variational parameters, prepared by shallower circuits and fewer gates compared to UCCSD~\cite{Grimsley2019ADAPT,qubitAdaptHoLun}. In this work, we introduce a pool-independent variation of the ADAPT-VQE algorithm, called TETRIS-ADAPT-VQE, in order to cut down on the resources needed to carry out a given simulation task. TETRIS-ADAPT-VQE achieves this in three ways:

\paragraph{Curtailing circuit depths:}

The original ADAPT-VQE algorithm adds a single operator per iteration, and more often than not, the sets of qubits that two consecutive operators act on intersect. That is, operators added in consecutive layers cannot be implemented simultaneously at the circuit level, but instead must be applied in succession. In contrast, TETRIS-ADAPT-VQE by design adds multiple operators in the same iteration, each acting on different sets of qubits, allowing them to be implemented simultaneously in the circuit. This results in significantly shallower circuits with roughly the same number of CNOT gates and variational parameters compared to the original ADAPT-VQE.

\paragraph{Slashing the measurement overhead:}

Although the operator gradient measurement step can in principle be performed in parallel, there will be a finite number of quantum processors available for a given experiment, and reducing the total number of shots is desirable. Because TETRIS-ADAPT-VQE adds multiple operators to the ansatz at each iteration while keeping their total number roughly constant, the number of times the pool gradient needs to be measured is only a fraction of that of the original algorithm. In the JW mapping, with the qubit and QE pools, the factor by which TETRIS-ADAPT-VQE reduces the gradient measurement overhead grows roughly linearly with the system size.

\paragraph{Exploring the Hilbert space faster:}

By adding multiple operators in tandem, TETRIS-ADAPT-VQE samples the Hilbert space faster and in more directions, especially in the early iterations. Because operator addition is still gradient-based and problem-aware, the ansatz does not leave the subspace where the solution lives. 

\section*{Acknowledgements}
This research was supported by the US Department of Energy (Award No. DE-
SC0019199). S.E.E. also acknowledges support from Award No. DE-SC0019318.


\providecommand{\noopsort}[1]{}\providecommand{\singleletter}[1]{#1}%

\end{document}